\documentclass[a4paper,UKenglish,cleveref, autoref, thm-restate]{lipics-v2021}

\pdfoutput=1 
\hideLIPIcs  
\usepackage{algorithm}
\usepackage{algpseudocode}
\usepackage{booktabs}
\usepackage{pifont}
\newcommand{\cmark}{\ding{51}}%
\newcommand{\xmark}{\ding{55}}%
\usepackage[disable]{todonotes} 

\title{A Neurosymbolic Approach to Loop Invariant Generation via Weakest Precondition Reasoning}

\titlerunning{A Neurosymbolic Approach to Loop Invariant Generation via WP Reasoning}

\author{Daragh King\footnote{Corresponding Author}}{Department of Computer Science and Statistics, Trinity College Dublin, Dublin, Ireland }{kingd6@tcd.ie}{}{}

\author{Vasileios Koutavas}{Lero, Research Ireland Centre for Software \and Department of Computer Science and Statistics, Trinity College Dublin, Dublin, Ireland }{vasileios.koutavas@tcd.ie}{}{}

\author{Laura Kov\'acs}{Faculty of Informatics, TU Wien, Vienna, Austria}{laura.kovacs@tuwien.ac.at}{}{}

\authorrunning{D. King, V. Koutavas, and L. Kov\'acs} 

\Copyright{Daragh King, Vasileios Koutavas, and Laura Kov\'acs} 

\ccsdesc[500]{Software and its engineering~Formal software verification}
\ccsdesc[500]{Theory of computation~Program verification}
\ccsdesc[300]{Computing methodologies~Artificial intelligence}

\keywords{Loop invariants, Program verification, Hoare logic, Neurosymbolic AI, Large language models}

\nolinenumbers 

\begin{document}

\maketitle

\begin{abstract}
Loop invariant generation remains a critical bottleneck in automated program verification. Recent work has begun to explore the use of Large Language Models (LLMs) in this area, yet these approaches tend to lack a reliable and structured methodology, with little reference to existing program verification theory. This paper presents NeuroInv, a neurosymbolic approach to loop invariant generation. NeuroInv comprises two key modules: (1) a neural reasoning module that leverages LLMs and Hoare logic to derive and refine candidate invariants via backward-chaining weakest precondition reasoning, and (2) a verification-guided symbolic module that iteratively repairs invariants using counterexamples from OpenJML. We evaluate NeuroInv on a comprehensive benchmark of 150 Java programs, encompassing single and multiple (sequential) loops, multiple arrays, random branching, and noisy code segments. NeuroInv achieves a $99.5\%$ success rate, substantially outperforming the other evaluated approaches. Additionally, we introduce a hard benchmark of $10$ larger multi-loop programs (with an average of $7$ loops each); NeuroInv’s performance in this setting demonstrates that it can scale to more complex verification scenarios.
\end{abstract}

\section{Introduction}
\label{sec:intro}
Loop invariant generation is a central challenge in automated program verification. Within Hoare logic~\cite{hoare-logic}, loop invariants are the inductive assertions for establishing the correctness of looping computations; loop invariants must: (a) hold before the loop, (b) be preserved by each loop iteration, and (c) imply the loop postcondition upon exit (see Section~\ref{hoare} for additional details on Hoare logic and loop invariants). Despite their conceptual simplicity, discovering suitable loop invariants is an undecidable problem and notoriously difficult to achieve with heuristics. Traditional static techniques—including abstract interpretation~\cite{cousot1977abstract}, logical abduction~\cite{calcagno2009bi}, symbolic execution~\cite{nguyen2017symlnfer}, and recurrence solving~\cite{humenberger2017invariant}—perform well in narrow domains but often struggle to generalise across programs with more complex structures, such as programs with multiple loops. Dynamic approaches such as~\cite{ernst2007daikon} can infer loop invariants from program executions, but their soundness is fundamentally constrained by the trace quantity and quality.

The advent of Large Language Models (LLMs) has led to recent work exploring whether they can generate program properties such as weakest preconditions~\cite{11024288, king2025fuzzfeedautomaticapproachweakest}, postconditions~\cite{endres2024can}, and even loop invariants~\cite{pei2023can, janssen2024can}. While these early investigations demonstrate promise, existing LLM-based approaches to loop invariant generation typically rely on ad-hoc prompting, and provide no systematic alignment with program verification theory. Moreover, because these methods lack a principled mechanism for the guided refinement of loop invariants, the invariants that are produced are often brittle 
(both syntactically and semantically) — these invariants may be sufficient for simple integer loops but unreliable for programs involving arrays, multiple loops, or noisy and semantically irrelevant code.

To address these limitations in LLM-based loop invariant inference, we introduce NeuroInv, a \emph{neurosymbolic}~\cite{garcez2023neurosymbolic} framework for loop invariant generation that integrates LLMs with weakest precondition (WP) reasoning~\cite{wp-calculus} and counterexample-guided symbolic verification. NeuroInv comprises two complementary modules. The neural module segments the input program into loop-free and looping components, it then leverages an LLM to perform three primary tasks: (1) calculate WPs of loop-free code segments, (2) generate candidate loop invariants, and (3) check and refine candidate loop invariants according to the preservation and exit obligations defined by Hoare logic (Section~\ref{hoare}) -- when the aforementioned obligations fail, the module attempts targeted refinement of the candidate loop invariant. The process employed by the neural module yields a structured and theory-guided reasoning procedure that stands in contrast to ad-hoc prompting approaches~\cite{janssen2024can}.

The symbolic module strengthens the above approach by validating the neurally-derived invariants using OpenJML, extracting counterexamples when verification fails, and attempting invariant repair through a feedback-directed process. This \emph{dual-level refinement and repair} approach enables NeuroInv to reliably generate correct loop invariants.

We evaluate NeuroInv on a comprehensive benchmark of 150 Java programs, encompassing single and multiple (sequential) loops, multiple arrays, random branching, and noisy code segments. NeuroInv achieves a $99.5\%$ success rate, substantially outperforming ad-hoc prompting, neural-only ablations, and a state-of-the-art LLM-based specification generator for Java~\cite{ma2024specgen}. A hard benchmark of $10$ larger multi-loop programs further demonstrates that NeuroInv scales effectively to more complex programs, with the neural module playing a key role in handling longer and more complex reasoning chains. Taken together, these results illustrate that systematic WP reasoning, when integrated with verification feedback, significantly enhances the robustness and reliability of LLM-assisted invariant generation.

The remainder of the paper is structured as follows. Section~\ref{hoare} outlines the requisite background on Hoare logic~\cite{Floyd67, hoare-logic}, weakest preconditions~\cite{wp-calculus}, and loop invariants necessary for understanding NeuroInv. Section~\ref{sec:method} presents the method behind NeuroInv, while Section~\ref{sec:evaluation} outlines our experimental evaluation and the obtained results. Section~\ref{sec:discussion} discusses threats to validity and future work. Section~\ref{sec:related} highlights related work, and Section~\ref{sec:conclusion} concludes.

\subsection{Hoare Logic, Weakest Preconditions and Loop Invariants}\label{hoare} 
Hoare logic provides a formal system for reasoning about the correctness of imperative programs. The core unit of this logic is the Hoare triple (Definition~\ref{HoareTriple}). Hoare triples fundamentally rely on the notions of pre- and postconditions: postconditions specify the desired property of the program state upon termination, whilst preconditions specify the required property of the initial program state for it to behave as intended (i.e.\ to satisfy the program's postcondition). The logic has a total and partial correctness version, depending on whether it proves termination or not. Here, we focus on partial correctness.

\begin{definition}[Hoare Triple (Partial Correctness)]\label{HoareTriple}
A Hoare triple $\{P\}\ C\ \{Q\}$ asserts that if the precondition $P$ holds before
executing command $C$, and $C$ terminates, then the postcondition $Q$ holds after
execution.
\end{definition}

Reasoning over Hoare triples enables compositional program verification, but to do so
effectively one must carefully consider the notion of a precondition. Consider the Hoare triple
$\{P\}\ C\ \{Q\}$, if $P$ is too restrictive (or \emph{strong}), then it may exclude
initial states that would in fact satisfy $Q$ after $C$ executes and terminates. It has been shown \cite{wp-calculus} that for a program $C$ and postcondition $Q$, there exists a unique largest precondition, called the \emph{weakest
precondition} (WP).

\begin{definition}[Weakest Precondition]
The weakest precondition $\text{WP}(C, Q)$ is the most general condition that, if true before executing command $C$, guarantees that postcondition $Q$ holds after execution (assuming termination).
\end{definition}

The WP calculus \cite{wp-calculus} provides a proof technique for Hoare Triples: to prove $\{P\}\ C\ \{Q\}$, one must compute $\text{WP}(C, Q)$, and prove that $P \Rightarrow \text{WP}(C, Q)$. This means that any starting state satisfying $P$ also satisfies $\text{WP}(C, Q)$, and that after executing $C$, the final state will satisfy $Q$.

\begin{definition}[Initialisation]\label{init-imp}
\begin{equation}
P \Rightarrow \text{WP}(C, Q)
\label{init-imp-env}
\end{equation}

\end{definition}

Crucially, the WP calculus offers compositional rules for computing weakest preconditions. A core rule of the calculus states that the WPs of sequential commands are composed in a backward manner: for a sequence of commands $C_1; C_2$, the WP of $C_2$ (with respect to an overall postcondition $Q$) serves as the immediate postcondition of $C_1$ (Definition~\ref{WP-sequences}).
(Other rules compose WPs in the case of branching and assignment commands.)

\begin{definition}[WP of Sequence]\label{WP-sequences}
\begin{equation}
\text{WP}(C_1; C_2, Q) = \text{WP}(C_1, \text{WP}(C_2, Q))
\end{equation}
\end{definition}

For loops however, calculating weakest preconditions is undecidable, and the WP calculus relies on externally provided \emph{loop invariants}.
Broadly speaking, loop invariants are assertions that hold before and after each loop iteration. In the context of the WP calculus however, sufficient loop invariants must also satisfy the following two conditions:

\begin{definition}[Preservation]\label{preservation-def}
If the invariant $I$ holds and the loop condition $B$ is true, executing the loop body $C$
must preserve the invariant. Equivalently:
\begin{equation}
I \land B \Rightarrow \text{WP}(C, I)
\end{equation}

\end{definition}

\begin{definition}[Exit]\label{exit-def}
When the invariant $I$ holds and the loop exits (i.e.\ $B$ is false), the loop's postcondition $Q$
must be satisfied.
\begin{equation}
I \land \neg B \Rightarrow Q
\end{equation}

\end{definition}

These conditions, together with equation (\ref{init-imp-env}) and the WP-composition rules, ensure that the WP calculus  correctly builds a verification proof within the framework of Hoare logic. In Section~\ref{neural-module}, the two conditions in Definitions~\ref{preservation-def} and~\ref{exit-def} play a major role in the loop invariant refinement process. Additionally, Definition~\ref{WP-sequences} forms the basis of the neural module's reasoning.

\section{Method}\label{sec:method}
\begin{figure}[t]
    \centering
    \includegraphics[width=0.90\linewidth]{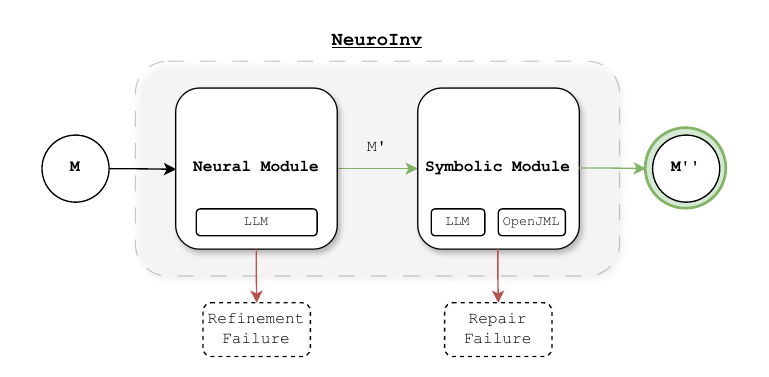}%
    \caption{High-level architecture of NeuroInv; $M$ is the input program, $M'$ is the input program with the neurally-derived loop invariants, and $M''$ represents $M$ with the correct loop invariants as JML annotations.} 
    \label{fig:arch}
\end{figure}

\begin{algorithm}[t]
\caption{Neurosymbolic Loop Invariant Generation}
\label{alg:main}
\footnotesize
\begin{algorithmic}[1]
\Require Java method $M$ with JML precondition $\text{Pre}$ and postcondition $\text{Post}$
\Ensure Verified method $M'$ with loop invariants

\State \textbf{Module 1: Neural Reasoning (Algorithm~\ref{alg:neural})}
\State $\text{invariants} \gets \textsc{NeuralReasoningModule}(M, \text{Pre}, \text{Post})$

\State
\State \textbf{Module 2: Symbolic Verification (Algorithm~\ref{alg:symbolic})}
\State $M' \gets \textsc{SymbolicVerificationModule}(M, \text{invariants}, $\text{Pre}$, $\text{Post}$)$

\State \Return $M'$
\end{algorithmic}
\end{algorithm}
Our approach, NeuroInv, aims to generate sufficient loop invariants for verifying the correctness of Java methods against their given pre- and postconditions. The system is built around a two-module neurosymbolic architecture that combines weakest precondition (WP) reasoning with symbolic verification.

\textbf{Neural Module:} Given a Java method with JML pre- and postconditions, the neural module attempts to derive high-quality candidate loop invariants via backward-chaining WP reasoning. The module leverages an LLM to perform three primary tasks: (1) calculate WPs of loop-free code segments; (2) generate candidate loop invariants; (3) check and refine candidate loop invariants according to the logical implications defined in Definitions~\ref{preservation-def} and~\ref{exit-def}. 

\textbf{Symbolic Module:} Given the candidate loop invariant(s) derived by the neural module, the symbolic module translates these into JML annotations and attempts program verification via OpenJML. If verification fails, the module extracts counterexample information from the OpenJML output, and attempts to use this to repair the candidate loop invariant. This process iterates until verification succeeds or a maximum repair limit is reached.

The key insight of our neurosymbolic design is \emph{dual-level refinement and repair}: The \emph{refinement} process employed by the neural module attempts to ensure that the candidate loop invariants are generated in a manner consistent with Hoare logic and the WP calculus. The \emph{repair} process of the symbolic module leverages verification feedback to repair loop invariants according to errors that have eluded the neural module. 

Figure~\ref{fig:arch} illustrates the high-level architecture of NeuroInv. Algorithm~\ref{alg:main} describes the overall approach, whilst Algorithms~\ref{alg:neural} and~\ref{alg:symbolic} depict the details of the neural and symbolic modules respectively. Additionally, Appendix \ref{prompts-appendix} contains NeuroInv's prompts.

\begin{figure}[t]
    \centering
    \includegraphics[width=1\linewidth]{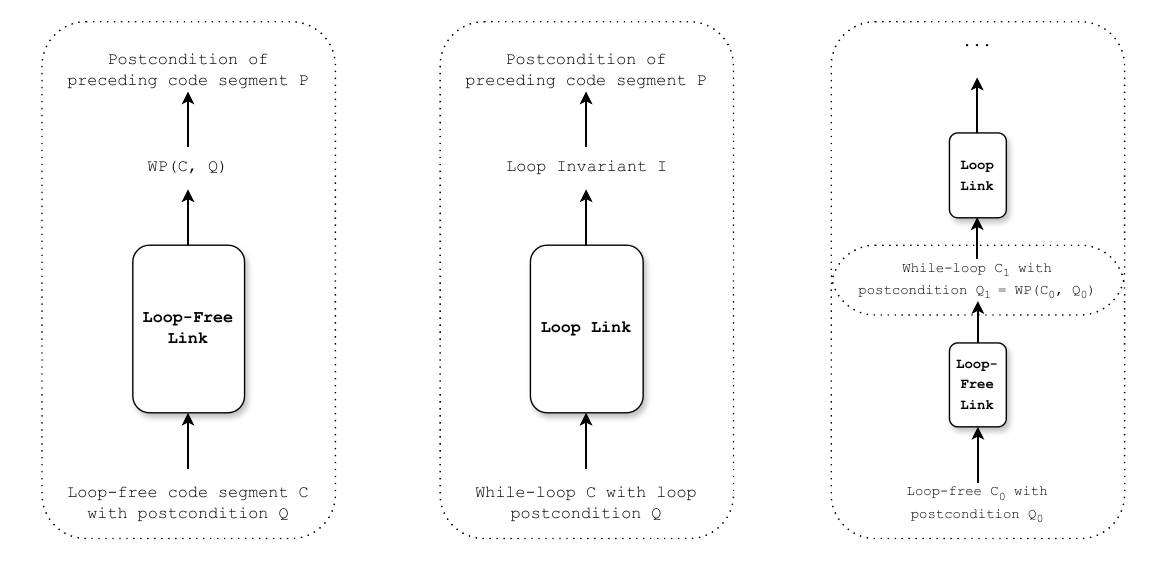}%
    \caption{The two left-most entities depict the link types used within the neural module's backward chaining WP reasoning. The entity on the right shows the chain between a "Loop-Free" and  "Loop" link, for a hypothetical program segment which is comprised of a \texttt{while} loop followed by loop-free code.} 
    \label{fig:links}
\end{figure}

\subsection{Neural Module}\label{neural-module}
The neural module generates candidate loop invariants via backward-chaining weakest precondition reasoning. This process relies on the LLM to perform three core tasks: (1) calculate WPs of loop-free code segments; (2) generate candidate loop invariants; (3) check and refine candidate loop invariants according to the logical implications of Definitions~\ref{preservation-def} and~\ref{exit-def}. This process follows Algorithm~\ref{alg:neural} below.

\begin{algorithm}[t]
\caption{Neural Module: Loop Invariant Generation via Backward-Chaining WP Reasoning}
\label{alg:neural}
\footnotesize
\begin{algorithmic}[1]
\Require Method $M$, precondition $\text{Pre}$, postcondition $\text{Post}$
\Ensure Candidate loop invariants, or ERROR if refinement limit is reached

\State \textbf{Phase 1: Program Segmentation}
\State $\langle \text{codeSegments}, \text{loops} \rangle \gets \textsc{SegmentProgram}(M)$

\State
\State \textbf{Phase 2: Backward-Chaining WP Reasoning}
\State $\text{Post} \gets \textsc{ExtractPostcondition}(M)$ \Comment{From JML postcondition}
\State $\text{currentPost} \gets \text{Post}$
\State $\text{invariants} \gets \{\}$

\If{$\exists$ final code segment after all loops}
    \State $\text{currentPost} \gets \textsc{LLM-ComputeWP}(\text{finalSegment}, \text{currentPost})$
\EndIf

\For{$\text{loopIndex} \gets |\text{loops}|$ \textbf{down to} $1$} \Comment{Process loops in reverse order}
    \State $\text{currentLoop} \gets \text{loops}[\text{loopIndex}]$
    \State $\text{loopCondition} \gets \text{condition}(\text{currentLoop})$
    \State $\text{loopBody} \gets \text{body}(\text{currentLoop})$
    \State $\text{loopPost} \gets \text{currentPost}$
    
    \State $\text{invariant} \gets \textsc{LLM-GenerateInvariant}(\text{currentLoop}, \text{loopPost})$
    \State $\text{validated} \gets \textbf{false}$
    
    \For{$\text{attempt} \gets 1$ \textbf{to} \text{MAX\_REFINEMENT}}
        \State $\text{weakestPre} \gets \textsc{LLM-ComputeWP}(\text{loopBody}, \text{invariant})$
        \State $\text{preservation} \gets \textsc{LLM-CheckImplication}(\text{invariant} \land \text{loopCondition} \Rightarrow \text{weakestPre})$
        \State $\text{exit} \gets \textsc{LLM-CheckImplication}(\text{invariant} \land \neg\text{loopCondition} \Rightarrow \text{loopPost})$
        
        \If{$\text{preservation} \land \text{exit}$}
            \State $\text{validated} \gets \textbf{true}$
            \State \textbf{break} \Comment{Both implications hold}
        \Else
            \State $\text{invariant} \gets \textsc{LLM-RefineInvariant}(\text{invariant}, \text{preservation}, \text{exit})$
        \EndIf
    \EndFor

    \If{$\neg \text{validated}$}
        \State \textbf{throw} \textsc{ERROR} \Comment{Refinement limit reached}
    \EndIf
    
    \State $\text{invariants}[\text{currentLoop}] \gets \text{invariant}$
    \State $\text{currentPost} \gets \text{invariant}$ \Comment{Invariant becomes postcondition for preceding code}
    
    \If{$\exists$ code segment before $\text{currentLoop}$}
        \State $\text{precedingSegment} \gets \text{codeSegments}[\text{loopIndex}]$
        \State $\text{currentPost} \gets \textsc{LLM-ComputeWP}(\text{precedingSegment}, \text{currentPost})$
    \EndIf
\EndFor

\State
\State \textbf{Precondition Sanity Check}
\State $\text{entailment} \gets \textsc{LLM-CheckImplication}(\text{Pre} \Rightarrow \text{currentPost})$ \Comment{Non-blocking}

\State \Return $\text{invariants}$
\end{algorithmic}
\end{algorithm}

\subsubsection{Program Segmentation}\label{segmentation}
Before attempting any reasoning, the neural module first splits the input program into looping and loop-free segments. Each segment is tagged with opening and closing marker comments, such as:
\begin{itemize}
    \item \texttt{// code N open} ... \texttt{// code N close}: The $N$th loop-free code segment.
    \item \texttt{// while N open} ... \texttt{// while N close}: The $N$th \texttt{while} loop
\end{itemize}
This segmentation enables systematic processing: the marker comments allow for the programmatic extraction of the desired code segments.

\subsubsection{Backward-Chaining WP Reasoning}
Starting from the program's given postcondition, the neural module works backwards through the segmented program (Section~\ref{segmentation}), chaining together WP calculations and loop invariant derivations. 

Conceptually, the aforementioned chains consist of two \textit{link} types, with each link type being associated to a particular type of code segment and set of LLM-based computations:
\begin{enumerate}
    \item \textbf{Loop-Free Link:}
    \begin{itemize}
        \item \textbf{Code segment type:} Loop-free.
        \item \textbf{Computation:} Calculating the WP of a loop-free code segment, given a postcondition $Q$.
    \end{itemize}
    \item \textbf{Loop Link:}
    \begin{itemize}
        \item \textbf{Code segment type:} \texttt{while} loop.
        \item \textbf{Computation (sequence):}
        \begin{enumerate}
            \item Generating a candidate loop invariant $I$, given a loop postcondition $Q$.
            \item Calculating the WP of the loop body $S$, using the loop invariant $I$ as the postcondition, i.e. calculating $WP(S, I)$.
            \item Verifying that $I$ satisfies Definition~\ref{preservation-def}.
            \item Verifying that $I$ satisfies Definition~\ref{exit-def}.
            \item Refining $I$ if steps $c$ or $d$ fail, if this occurs, then the sequence restarts at step $b$ with the refined loop invariant $I$.
        \end{enumerate}
    \end{itemize}
\end{enumerate}

Fig.~\ref{fig:links} depicts the two link types; the ultimate goal of both links is to produce the postcondition for the preceding code segment (if it exists). By chaining together these links, the neural module can work backwards from a program's given postcondition to derive loop invariants. At the end of this process, the neural module checks that last WP derived satisfies Definition~\ref{init-imp}.

\subsection{Symbolic Module}\label{symbolic-module}
The symbolic module is conceptually simple and can be summarised by the following steps:
\begin{enumerate}
    \item Translating loop invariants produced by the neural module into the corresponding JML annotations.
    \item Attempting to verify the annotated program using OpenJML.
    \item If Step~2 fails, extracting counterexample information from OpenJML and using it to repair the loop invariant(s). The process then repeats from Step~2 with the repaired invariant(s), until either (a) verification succeeds or (b) the maximum repair limit is reached.
\end{enumerate}
Algorithm~\ref{alg:symbolic} details this procedure.

\begin{algorithm}[t]
\caption{Symbolic Module: Verification with Counterexample-Guided Repair}
\label{alg:symbolic}
\footnotesize
\begin{algorithmic}[1]
\Require Method $M$, candidate invariants $\text{invs}$, precondition $\text{Pre}$, postcondition $\text{Post}$
\Ensure Verified method $M'$ or ERROR

\State \textbf{Phase 3: Finalization and Verification}
\State $M' \gets \textsc{LLM-ConvertToJML}(M, \text{invs})$ \Comment{Convert to JML syntax}

\For{$\text{repairAttempt} \gets 1$ \textbf{to} \text{MAX\_REPAIR}}
    \State $\langle \text{verified}, \text{counterexamples} \rangle \gets \textsc{OpenJML-Verify}(M')$
    
    \If{$\text{verified}$}
        \State \Return $M'$ \Comment{Successfully verified}
    \EndIf
    
    \State $\text{invs} \gets \textsc{LLM-RepairInvariant}(\text{invs}, \text{counterexamples})$
    \State $M' \gets \textsc{LLM-ConvertToJML}(M, \text{invs})$
\EndFor

\State \Return \text{ERROR} \Comment{Could not verify within the repair limit}
\end{algorithmic}
\end{algorithm}

\section{Evaluation}
\label{sec:evaluation}

\subsection{Experimental Setup}\label{experimental-setup}
The following sections outline the basic components of our experimental evaluation. Section~\ref{benchmark-suite} describes the the benchmark suite. Section~\ref{experimental-methods} describes the methods evaluated. Section~\ref{implementation-details} presents the implementation details. And finally, Section~\ref{evaluation-metrics} details the evaluation metrics used.

\subsubsection{Benchmark Suite}\label{benchmark-suite}
The benchmark suite used in our evaluation is divided into two subsets: (1) the standard benchmark and (2) the hard benchmark. The programs of these subsets share common attributes, each features: (a) single Java methods annotated with JML pre- and postconditions, and (b) at least one \texttt{while} loop.
\footnote{Both standard and hard benchmarks will be archived in Zenodo.}

\textbf{Standard Benchmark.} 
The standard benchmark consists of 150 programs, with the following split into high-level categories (summarised in Table~\ref{tab:standard-benchmark}):
\begin{enumerate}
    \item $62$ programs are single-loop Java translations of programs originating from the Code2Inv~\cite{si2020code2inv} dataset; these $62$ programs were chosen because they do not feature non-determinism in their loop guards.
    \item $35$ programs are array-manipulating single-loop programs derived from the work of~\cite{king2025fuzzfeedautomaticapproachweakest}; these $35$ programs can feature multiple arrays, and require loop invariants with universal quantification.
    \item $21$ programs are obtained as semantics-preserving multi-loop decompositions of a subset of the programs described in point~2 above.
    \item $15$ programs are obtained as semantics-preserving "noisy" translations of a subset of the programs described in points 1 and 2 above. These programs feature dead, irrelevant, or unreachable code, alongside (potentially) misrepresentative variable names.
    \item $10$ programs are single-loop Java translations of programs originating from~\cite{11024288}; each of these programs feature random branching within their respective loops.
    \item $5$ programs are array-manipulating single-loop programs derived from the work of~\cite{king2025fuzzfeedautomaticapproachweakest}; these $5$ programs can feature multiple arrays, and require loop invariants with existential quantification.
    \item $2$ programs fall outside of the above categories, one program creates a reversal of a given array, and the other checks for equality between two arrays.
\end{enumerate}

\begin{table}[t]
\centering
\caption{Standard benchmark composition and key characteristics.}
\label{tab:standard-benchmark}
\begin{tabular}{c r c c c c c c}
\toprule
Category & Count &
Arrays &
Multi-loop &
Noisy &
Rand. Branch &
$\forall$-Inv &
$\exists$-Inv \\
\midrule
1.             & $62$ & \xmark & \xmark & \xmark & \xmark & \xmark & \xmark \\
2. & $35$ & \cmark & \xmark & \xmark & \xmark & \cmark & \xmark \\
3.                     & $21$ & \cmark & \cmark & \xmark & \xmark & \cmark & \xmark \\
4.               & $15$ & \cmark$^{\dagger}$ & \xmark & \cmark & \xmark & \cmark$^{\dagger}$ & \xmark \\
5.           & $10$ & \cmark & \xmark & \xmark & \cmark & \cmark & \xmark \\
6. & $5$ & \cmark & \xmark & \xmark & \xmark & \xmark & \cmark \\
7.                         & $2$ & \cmark & \xmark & \xmark & \xmark & \cmark & \xmark \\
\midrule
Total                                       & $150$ & & & & & & \\
\bottomrule
\end{tabular}

\vspace{0.5ex}
\raggedright
\footnotesize
$^{\dagger}$Only a subset of the noisy programs are array-based and require universally quantified invariants (those derived from Category 2.).
\end{table}

\textbf{Hard Benchmark.} The hard benchmark features $10$ programs which are deliberately designed to stress-test the evaluated methods (Section~\ref{experimental-methods}) and their scalability. The characteristics of this benchmark are described in Section~\ref{hard-results}.

\subsubsection{Methods Evaluated}\label{experimental-methods}
We evaluate four methodologies in our experiments: ad-hoc prompting, NeuroInv (Section~\ref{sec:method}), NeuroInv*, and SpecGen~\cite{ma2024specgen}.

\textbf{Ad-hoc prompting:} this approach directly asks the LLM for the correct loop invariants for a given program; this is done without providing any annotated examples, and can thus be viewed as a zero-shot prompting approach~\cite{schulhoff2024prompt}. The prompt utilised can be seen in Listing~\ref{lst:zero-prompt-loop-inv} of Appendix~\ref{prompts-appendix}.

\textbf{NeuroInv:} this is our proposed neurosymbolic approach to loop invariant generation detailed in Section~\ref{sec:method}.

\textbf{NeuroInv*:} this approach consists only of the neural module outlined in Section~\ref{neural-module}. We include this “neural-only” variant as an ablation to better understand the contributions of the neural and symbolic modules. Note that NeuroInv subsumes NeuroInv*, therefore both approaches will share certain results in the following sections.

\textbf{SpecGen:}
SpecGen~\cite{ma2024specgen} is a recent LLM-based approach to automatically generating full program specifications (including loop invariants and pre- and postconditions). Consequently, the intended goals of SpecGen are broader than loop invariant generation alone -- SpecGen may also alter the given JML pre- and postconditions, since it was originally designed for use with unspecified programs (experimentally, we found that this occurred for approximately $68\%$ of programs in the standard benchmark). In our setting, we regard such cases as failures in the statistics reported in Section~\ref{main-results}, this is because we can no longer guarantee that the verified pre- and postconditions are equivalent to the originals. Despite this, we include SpecGen in our evaluation as it is the most closely related approach to our own: SpecGen targets Java, produces JML-based specifications, and can generate loop invariants -- to the best of our knowledge, a more suitable LLM-based loop invariant generator for Java does not exist currently. Moreover, SpecGen has been shown to outperform other related methods for specification generation, such as AutoSpec~\cite{wen2024enchanting}, Daikon~\cite{ernst2007daikon}, and Houdini~\cite{flanagan2001houdini}.

\subsubsection{Implementation Details}\label{implementation-details}
We utilise the following configuration for our experiments.
\begin{itemize}
    \item \textbf{Experimental Protocol:} We complete $5$ independent experimental runs of the methods outlined in Section~\ref{experimental-methods}, this is done for both the standard and hard benchmark evaluations.
    \item \textbf{Model:} We use \texttt{o4-mini}~\cite{o4-mini-ref} as the underlying model for each method evaluated (including SpecGen). At the time of writing, \texttt{o4-mini} is the most performant and affordable of OpenAI's reasoning models.
    \item \textbf{OpenJML and Solver:} we use \texttt{openjml 21-0.11} and \texttt{Z3 version 4.13.4 - 64 bit} as its underlying solver.
    \item \textbf{Refinement and Repair Parameters:} 
    \begin{enumerate}
        \item \texttt{MAX\_REFINEMENT:} refers to the maximum number of refinement attempts that can be made by the neural module for each loop in a given program. We use \texttt{MAX\_REFINEMENT}$=5$.
        \item \texttt{MAX\_REPAIR:} refers to the maximum number of repair attempts that can be made by the symbolic module per program. We use \texttt{MAX\_REPAIR}$=5$
    \end{enumerate}
\end{itemize}

\subsubsection{Evaluation Metrics}\label{evaluation-metrics}
The following metrics form the basis of our analysis.
\begin{itemize}
    \item \textbf{Success Rate:} this denotes the percentage of programs for which the generated loop invariants enable successful verification. The "Avg." success rate in Table~\ref{tab:standard-results-overview} denotes the mean success rates obtained over 5 experimental runs of all programs in the standard benchmark (i.e. 750 total program evaluations).
    \item \textbf{Standard Deviation:} we use this to measure the stability of the "Success Rate" results across multiple experimental runs.
    \item \textbf{Pass@$\mathit{k}$:} this metric quantifies the probability that at least one solution (out of $k$) is correct for a given problem~\cite{chen2021evaluating}. We use this metric on a per-program basis to observe how increasing $k$ affects the probability that at least one of the \textit{k} generated loop invariants (or loop invariant sets, in the multi-loop case) for a program will enable successful verification. The "Avg." pass@k values denoted in Table~\ref{tab:standard-results-overview} represent the mean pass@3 and pass@5 values obtained for the standard benchmark.
    \item \textbf{Refinement Iterations:} this denotes the average number of attempts the neural module has made to refine candidate loop invariants per program -- this is computed across a set of programs, e.g. the standard or hard benchmark. The "Avg." Refinement Iterations in Table~\ref{tab:standard-results-overview} denotes the mean number of Refinement Iterations observed across 5 experimental runs on the standard benchmark.
    \item \textbf{Repair Iterations:} this denotes the average number of attempts the symbolic module has made to repair loop invariant(s) per program -- this is computed across a set of programs, e.g. the standard or hard benchmark. The "Avg." Repair Iterations in Table~\ref{tab:standard-results-overview} denotes the mean number of Repair Iterations observed across 5 experimental runs on the standard benchmark.
\end{itemize}
\subsection{Main Results}\label{main-results}

\begin{table}[t]
\centering
\caption{Overall standard benchmark results, obtained across 5 experimental runs.}
\label{tab:standard-results-overview}
\begin{tabular}{lcccccc}
\toprule
Method
& \begin{tabular}{@{}c@{}}Avg. Success Rate\\(\%)\end{tabular}
& \begin{tabular}{@{}c@{}}Std. Dev.\\(\%)\end{tabular}
& \begin{tabular}{@{}c@{}}Avg. Refinement\\Iters.\end{tabular}
& \begin{tabular}{@{}c@{}}Avg. Repair\\Iters.\end{tabular}
& \begin{tabular}{@{}c@{}}Avg.\\Pass@3\end{tabular}
& \begin{tabular}{@{}c@{}}Avg.\\Pass@5\end{tabular} \\
\midrule
Ad-hoc           & 66.4 & 3.11 & -   & -   & 0.87   & 0.91  \\
SpecGen$^{\dagger}$ & 28.5 & 2.47    & -   & -   &  - & -   \\
NeuroInv*        & 84.7 & 2.31 &   0.59  & -    & 0.95   & 0.97   \\
NeuroInv         & 99.5 & 0.87 & 0.59     &  0.24   & 1  & 1  \\
\bottomrule
\end{tabular}
\vspace{1ex}
\raggedright
\footnotesize
$^{\dagger}$See Section~\ref{experimental-methods} for SpecGen caveats.

\end{table}

\subsubsection{Overall Performance}\label{overall-performance}
\begin{figure}[t]
    \centering
    \includegraphics[width=0.75\linewidth]{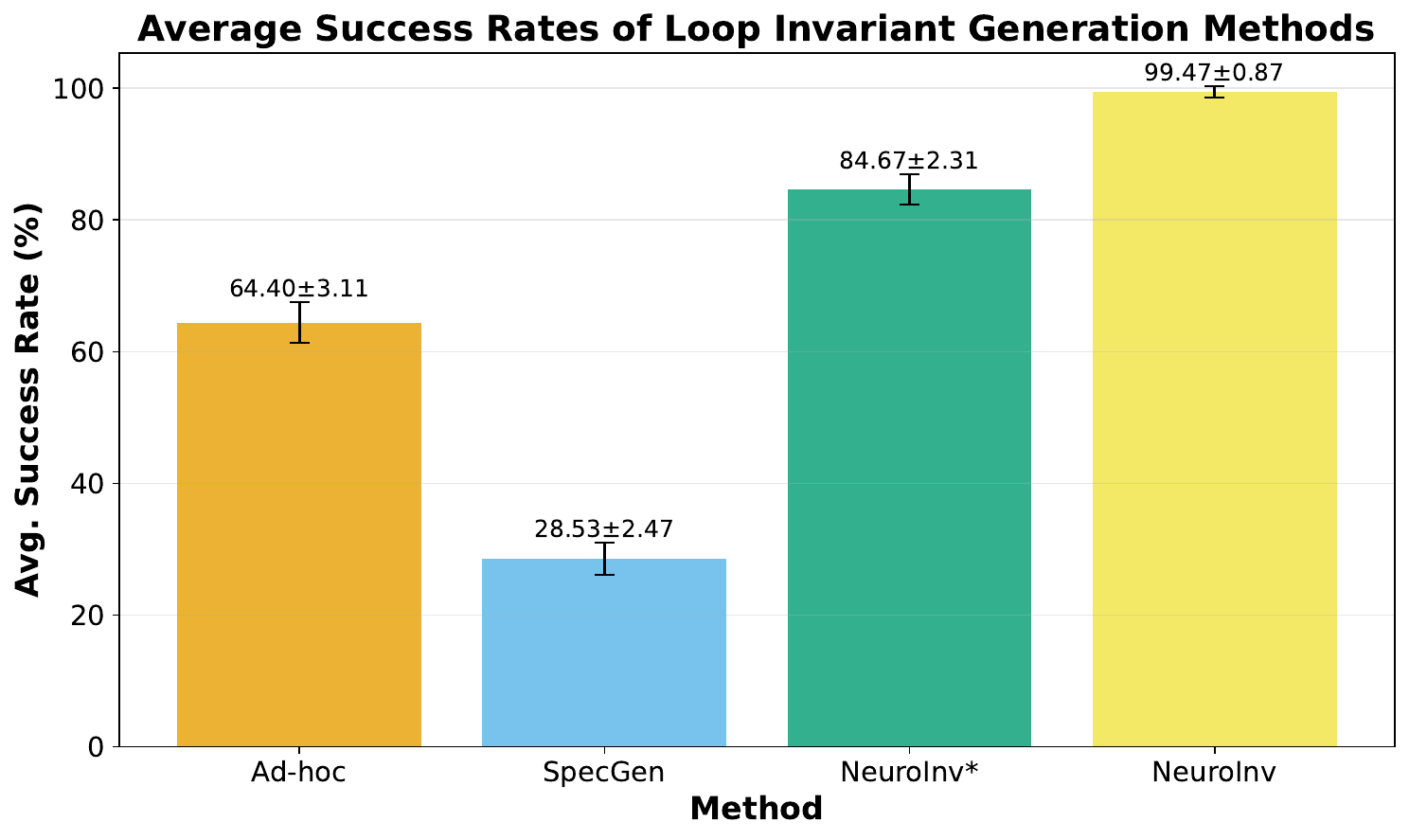}%
    \caption{Average success rates (with std. deviation bars), obtained across 5 experimental runs.}
    \label{fig:standard-success-rates}
\end{figure}

Table~\ref{tab:standard-results-overview} and Fig.~\ref{fig:standard-success-rates} present the overall results obtained on the standard benchmark. 

In terms of the average success rate observed, NeuroInv clearly trumps the other approaches; with a value of $99.5\%$, NeuroInv far surpasses the performance of both SpecGen ($28.5\%$) and ad-hoc prompting ($66.4\%$). Although NeuroInv* ($84.7\%$) fares better than both SpecGen and ad-hoc prompting, it falls short of the full NeuroInv approach. This suggests that the addition of the symbolic module enables the generation of higher quality loop invariants, i.e. the counter-example feedback provided by OpenJML can practically aid the repair of candidate loop invariants. It is worth noting that the relatively poor performance of SpecGen is likely due to it being an approach for generating full program specifications, and not loop invariants specifically. The performance gap between NeuroInv and SpecGen suggests that a blanket approach to program specification will not suffice for generating intricate loop invariants in the presence of given pre- and postconditions.

Considering result stability, NeuroInv achieves the lowest standard deviation of the success rate. Taken together with its superior average success rate, these two measures show that NeuroInv can consistently generate correct loop invariants -- this notion is further supported by NeuroInv’s average pass@$3$ score being equal to $1$ (this means that in any given 3 experimental runs, NeuroInv finds the correct loop invariant(s) for any given standard benchmark program at least once). Although NeuroInv* is less stable than NeuroInv, its standard deviation is still lower than both ad-hoc prompting and SpecGen; suggesting that the systematic approach employed by the neural module leads to correct loop invariants consistently.

\medskip\noindent
\textbf{Takeaway:} On our benchmarks, NeuroInv reliably generates the most correct loop invariants ($99.5\%$); SpecGen fares the worst ($28.5\%$), and is even succeeded by an ad-hoc prompting approach ($66.4\%$) specific to loop invariant generation.

\subsubsection{Refinement Results}\label{refinement-results}
In this section we present the loop invariant refinement results derived by NeuroInv's neural module (i.e. NeuroInv*) on the standard benchmark.

Recall that a loop invariant refinement occurs in the neural module (Section~\ref{neural-module}) when the LLM deems that a candidate invariant does not satisfy at least one of Definition~\ref{preservation-def} (i.e. an Implication 1 failure) and Definition~\ref{exit-def} (i.e. an Implication 2 failure). Note that an Implication 1 failure precludes the possibility of an Implication 2 failure being recorded -- this is a by-product of the current implementation.

\begin{figure}[t]
    \centering
    \includegraphics[width=0.75\linewidth]{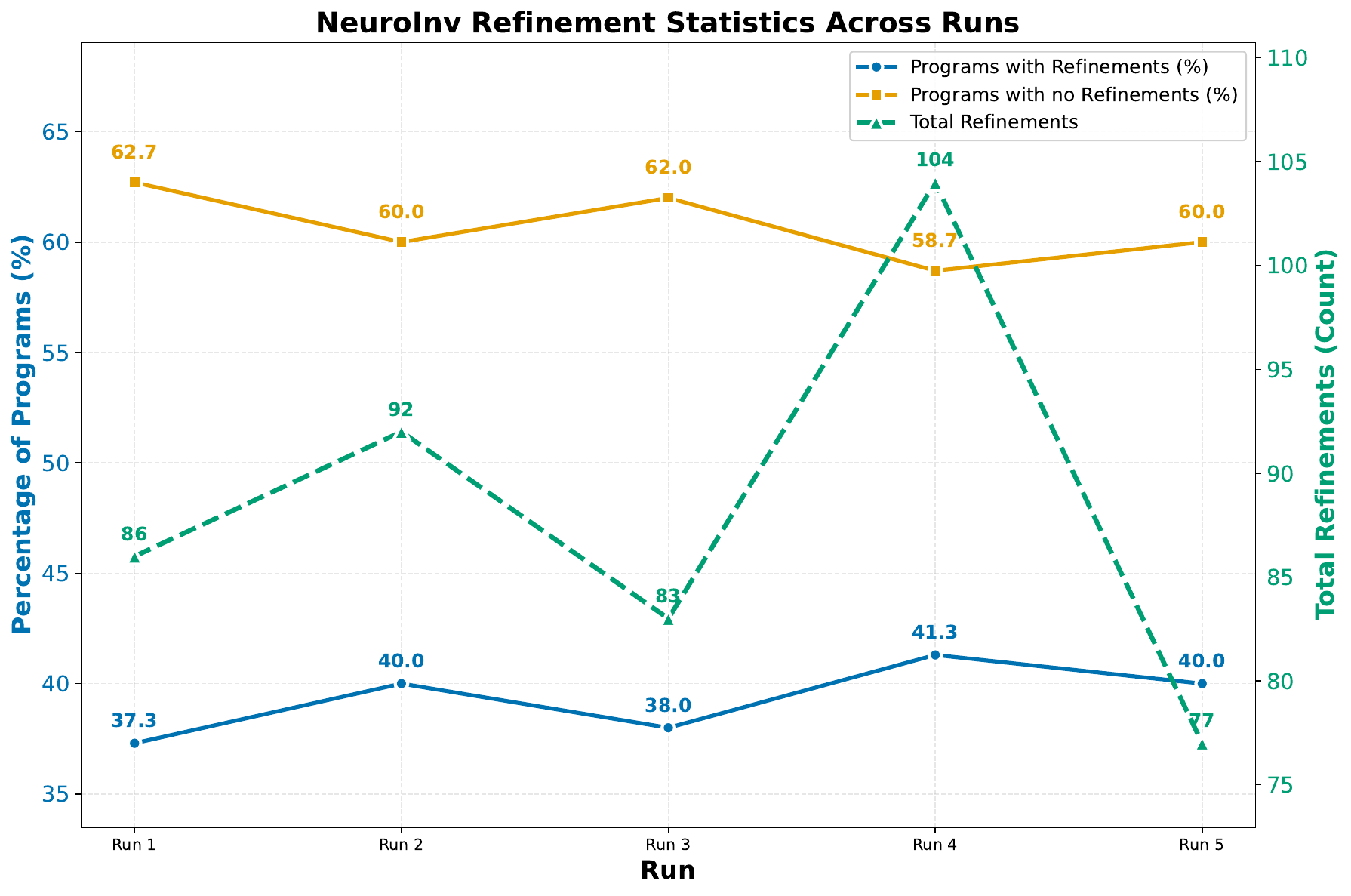}%
    \caption{The proportions of programs which do (and do not) require loop invariant refinements across each experimental run (solid lines). Additionally, the dashed line depicts the total number of refinements required per experimental run.} 
    \label{fig:multi_line_trends}
\end{figure}

Fig.~\ref{fig:multi_line_trends} depicts the proportion of programs which require (and do not require) loop invariant refinements; the stability of these plots suggests that the number of programs requiring loop invariant refinements is relatively consistent. This consistency is somewhat surprising when considering the "Total Refinements" plot (i.e. the dashed-line); with minimum and maximum values of $77$ and $102$ respectively, this statistic appears to be more erratic in comparison. Taken together, these observations suggest that there is not a strong relationship between the number of programs requiring refinements and the total number of refinements entailed.

\begin{figure}[t]
    \centering
    \includegraphics[width=0.75\linewidth]{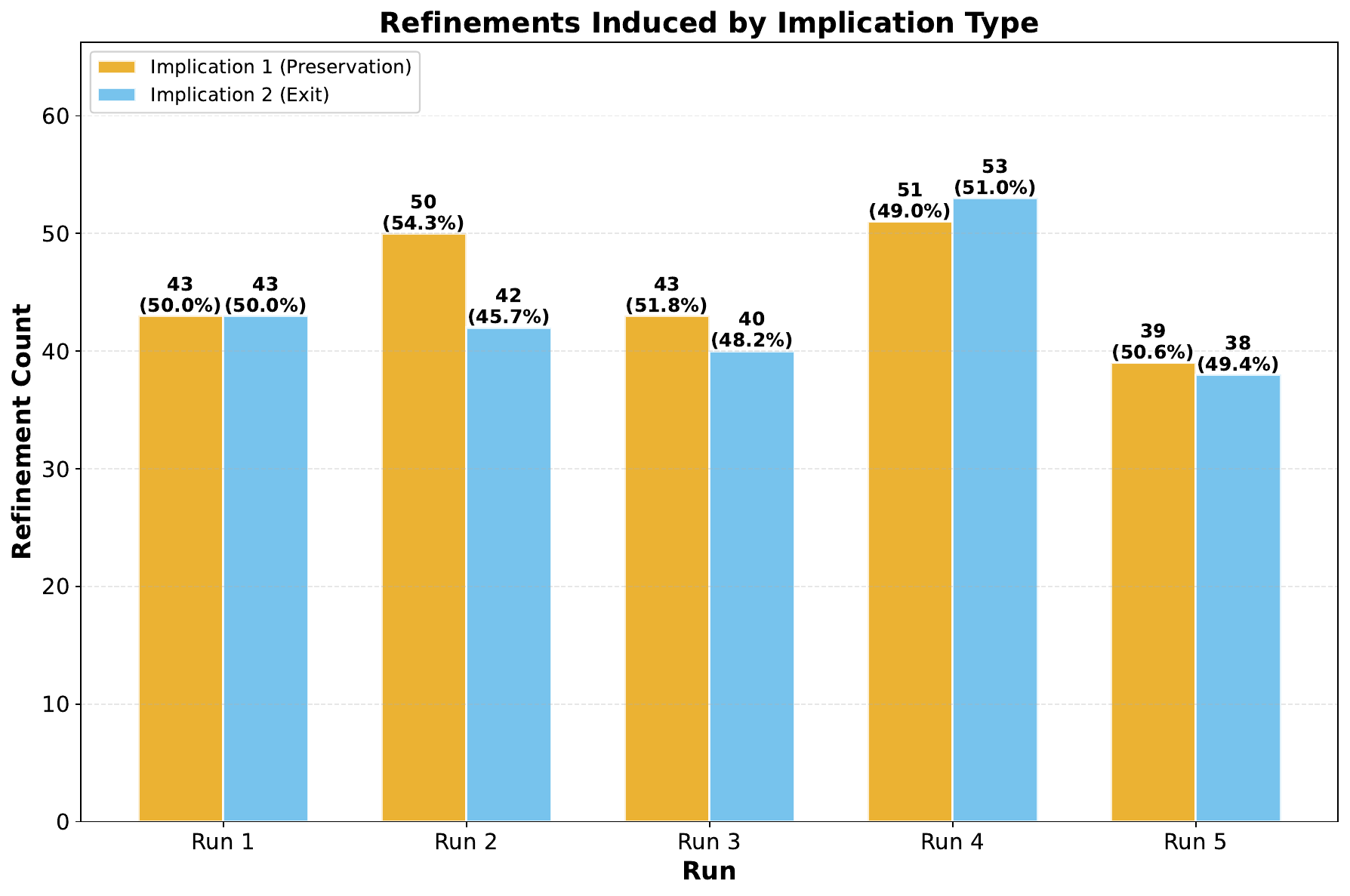}%
    \caption{The proportions of refinements induced by Implication 1 and Implication 2 failures, evaluated across 5 experimental runs.}
    \label{fig:standard-implication-failures}
\end{figure}

Fig.~\ref{fig:standard-implication-failures} displays the proportion of refinements induced by Implication 1 and Implication 2 failures; the split between the two failure types is close to $50:50$ for each experimental run.

\medskip\noindent
\textbf{Takeaway:} A consistently-sized subset of programs trigger loop invariant refinement; the type of logical failure which triggers refinement is evenly split between Implication 1 and Implication 2 failures; and the number of refinements required varies considerably across experimental runs.



\subsubsection{Repair Results and Ablation Analysis}\label{ablation-symbolic}
In order to understand how the individual modules of our neurosymbolic approach contribute to producing correct loop invariants, we further discuss NeuroInv* (i.e. the neural module) and NeuroInv (the combined neural and symbolic modules).

The overall standard benchmark results (Table~\ref{tab:standard-results-overview}) show that NeuroInv* and NeuroInv achieve average success rates of $84.7\%$ and $99.5\%$ respectively. In other words, the inclusion of the symbolic module leads to an average success rate increase of $+14.8\%$. On the other hand, recall that loop invariant repair is carried out by the symbolic module (Section~\ref{symbolic-module}) when the loop invariants produced by the neural module (i.e. NeuroInv*) do not enable program verification with OpenJML. Considering NeuroInv*'s average success rate of $84.7\%$ (Table~\ref{tab:standard-results-overview}), this implies that the vast majority of cases never require invariant repair by the symbolic module. This notion is further backed by the average repair iterations statistic ($0.24)$ denoted in Table~\ref{tab:standard-results-overview}.

Additionally, the pass@$5$ scores in Table~\ref{tab:standard-results-overview} show that there exists a subset of programs for which NeuroInv* never derives the correct loop invariant(s), whilst NeuroInv does at least once. We will refer to this set of programs as the "problem" set.

Upon investigation, we find that the problem set contains $5$ programs and showcases a key limitation of the neural-only approach. For each program in the problem set, the candidate loop invariant generated by the neural module failed OpenJML verification -- in each case this was due to the invariant not holding before the loop. In fact, the neural module was also able to highlight this limitation, this was done by checking the implication of Definition~\ref{init-imp} at the end of its reasoning process. At this stage though, it is simply too little, too late -- the current implementation of the neural module has no effective recourse to backtrack along its reasoning process at this point -- hence, the feedback provided by OpenJML and the symbolic module is \emph{essential} here for generating the correct loop invariants.

\medskip\noindent
\textbf{Takeaway:} NeuroInv's inclusion of the symbolic module leads to an average success rate increase of $+14.8\%$ over NeuroInv*, and the symbolic module is essential to producing the correct loop invariants for a small subset of standard benchmarks.

\subsection{A Brief Scalability Analysis: The Hard Benchmark}\label{hard-results}
\begin{table}[t]
\centering
\caption{Hard benchmark composition and key characteristics.}
\label{tab:test-difficult}
\begin{tabular}{crrrccr}
\toprule
Program & \# Loops & \# Arrays & $\forall$-Inv & $\exists$-Inv & Noisy & Lines Of Code \\
\midrule
1 & 2 & 0 & \xmark & \xmark & \xmark & 29 \\
2 & 3 & 0 & \xmark & \xmark & \xmark & 43 \\
3 & 4 & 0 & \xmark & \xmark & \xmark & 56 \\
4 & 5 & 0 & \xmark & \xmark & \xmark & 77 \\
5 & 4 & 2 & \checkmark & \xmark & \xmark & 62 \\
6 & 4 & 3 & \checkmark & \checkmark & \xmark & 62 \\
7 & 8 & 1 & \checkmark & \xmark & \xmark & 78 \\
8 & 8 & 3 & \checkmark & \checkmark & \xmark & 95 \\
9 & 10 & 2 & \checkmark & \xmark & \checkmark & 152 \\
10 & 12 & 2 & \checkmark & \xmark & \xmark & 117 \\
\bottomrule
\end{tabular}
\end{table}
In order to test the scalability of NeuroInv to complex programs, we created a hard benchmark consisting of $10$ programs; Table~\ref{tab:test-difficult} depicts the characteristics of these programs. Following the protocol of the standard benchmark evaluation (i.e. using the implementation details of Section~\ref{implementation-details}), we conduct 5 experimental runs of ad-hoc prompting (Section~\ref{experimental-methods}) and NeuroInv (and its subsumed NeuroInv*) -- we refrain from evaluating SpecGen in this context due to its performance on the standard benchmark (Section \ref{tab:standard-results-overview}). Table~\ref{tab:hard-results-overview} presents these results.

NeuroInv achieves an average success rate of $62\%$, whilst ad-hoc prompting and NeuroInv* only achieve $14\%$ and $40\%$ respectively  -- NeuroInv succeeds ad-hoc prompting by a large margin of $+48\%$ points. These results show that NeuroInv can scale to difficult programs (albeit with a hit in performance when compared to standard benchmark programs), whilst ad-hoc prompting fails to do so. These results also highlight that there is still value in the systematic approach employed by NeuroInv* (i.e. NeuroInv's neural module) in this setting. 

Additionally, the neural module clearly appears to be performing more refinements; the average value of $2.9$ in Table~\ref{tab:hard-results-overview} is an almost $5$x increase when compared to the corresponding value for the standard benchmark ($0.59$) -- this is likely due to the large increase in the average number of loops per program in the hard benchmark. In contrast, the average number of repairs being carried out by the symbolic module only increased by approximately $2$x ($0.54$ vs. $0.24)$; upon investigation, we found that the vast majority of failures ($94\%$) occurred whilst the neural module was performing invariant refinements (i.e. the max number of refinements for a particular loop was reached and the process was aborted) -- therefore, the symbolic module had fewer chances to perform invariant repair.

\medskip\noindent
\textbf{Takeaway:} NeuroInv scales to complex programs and succeeds ad-hoc prompting by a significant margin ($+48\%$ points); the application of NeuroInv to hard benchmark programs results in more loop invariant refinements being performed by the neural module, and the vast majority of observed failures occur during this process ($94\%$).

\begin{table}[t]
\centering
\caption{Overall hard benchmark results, obtained across 5 experimental runs.}
\label{tab:hard-results-overview}
\begin{tabular}{lcccc}
\toprule
Method
& \begin{tabular}{@{}c@{}}Avg. Success Rate\\(\%)\end{tabular}
& \begin{tabular}{@{}c@{}}Std. Dev.\\(\%)\end{tabular}
& \begin{tabular}{@{}c@{}}Avg. Refinement\\Iters.\end{tabular}
& \begin{tabular}{@{}c@{}}Avg. Repair\\Iters.\end{tabular}\\
\midrule
Ad-hoc           & 14 & 5.48 & -   & -  \\
NeuroInv*        & 40 & 7.07 &   2.9  & -       \\
NeuroInv         & 62 & 10.95 & 2.9     &  0.54     \\
\bottomrule
\end{tabular}
\vspace{1ex}
\raggedright
\footnotesize

\end{table}


\section{Discussion}
\label{sec:discussion}

\subsection{Threats to Validity}
\label{sec:validity-threats}
The primary threats to the validity of this work stem from our usage of LLMs. The outputs of LLMs are inherently stochastic, hence the results of this study are not exactly reproducible -- we attempt to mitigate this threat by performing multiple experimental runs for each evaluation. Data leakage poses a similar threat -- data leakage occurs when the information from a test set is present in the training data of the LLM. We attempt to alleviate this particular threat by deriving a large proportion of our benchmark programs from works after the knowledge cut off of \texttt{o4-mini} (i.e. after the $1$st of June  $2024$); additionally, the benchmark programs used with origins prior to this date are translations from their original programming language (e.g. from C to Java for the Code2Inv derived benchmark programs, Section~\ref{benchmark-suite}). Another threat arises when considering this work's sole usage of \texttt{o4-mini}; the argument could be made that the results obtained are biased in this respect, and that other models, from other vendors, may perform significantly better or worse when used in tandem with the methods evaluated (Section~\ref{experimental-methods}).

\subsection{Future Work}
\label{sec:future}
Several avenues of future work arise naturally from this study.

The links (introduced in Section~\ref{neural-module}) which constitute the backward-chaining WP reasoning of the neural module are currently \emph{fragile}. In other words, there are currently no strong guarantees about what these links produce; for example, the weakest preconditions produced by the loop-free link type (Fig.~\ref{fig:links}) are not formally verified in any capacity -- we assume that if the final loop invariant(s) for a given program are correct, then the constituent WPs are likely correct also. This use of unverified LLM-generated WPs is not without basis, prior works such as~\cite{11024288} and~\cite{king2025fuzzfeedautomaticapproachweakest} show that reasoning models can often generate correct method-level WPs without additional tool intervention or feedback. Alleviating the fragility of these links by incorporating a verifier earlier in the invariant generation process will allow for potential errors to become localised; we leave this intervention to future work.

In a similar vein, Section~\ref{ablation-symbolic} highlights an inherent flaw of the neural module. The current implementation has no recourse to backtrack along its reasoning process when it finds that the final generated weakest precondition for a program does not satisfy Definition~\ref{init-imp}. Implementing this ability to effectively backtrack is a valuable line of future work; the results of Section~\ref{ablation-symbolic} suggest that a solution to this issue will enable the neural module to effectively generate more correct loop invariants without need for feedback from OpenJML.

The expansion and standardisation of the hard benchmark (Section~\ref{hard-results}) is another relevant line of future work; to the best of our knowledge, a similar such benchmark of difficult (i.e. containing $2+$ loops over multiple arrays, etc.) Java-based loop invariant generation problems does not exist.

\section{Related Work}
\label{sec:related}
Loop invariant generation is a central challenge in automated program verification. While no complete solution can exist due to theoretical limitations, substantial prior work has approached the problem through a variety of methods.

\subsection{Traditional Techniques}
We define traditional techniques as those that do not employ a learning-based approach or rely on LLMs; these techniques span the spectrum of dynamic and static program analysis.

Dynamic approaches attempt to infer loop invariants from program execution traces. Daikon~\cite{ernst2007daikon}, a popular invariant generation tool for Java, attempts to derive invariants (including pre- and postconditions) for a given program from its test suite traces. Despite its capabilities, Daikon is innately constrained by the quality and coverage of the given test-suite; as a  consequence, the invariants that Daikon generates may not hold universally. 

On the other hand, static approaches attempt to reason about loop invariants without program execution; these techniques typically rely upon strong theoretical foundations and make stronger guarantees when compared to dynamic approaches. A wide array of techniques have evolved in this space and found application towards loop invariant generation, e.g. abstract interpretation~\cite{cousot1977abstract}, logical abduction~\cite{calcagno2009bi}, recurrence analysis~\cite{humenberger2017invariant}, symbolic execution~\cite{nguyen2017symlnfer}, etc. These techniques tend to excel within their well-defined contexts, but fail to scale and generalise more broadly.

\subsection{Learning- and LLM-Based Approaches}
Learning-based approaches utilise machine learning in some capacity to aid loop invariant generation. Code2Inv~\cite{si2020code2inv} is a prime example of such an approach: it uses deep reinforcement learning (RL) to learn loop invariants through interactions with a proof checker. In a similar vein, LIPuS~\cite{yu2023loop} uses RL-based pruning and SMT solving to generate loop invariants.

More recently, researchers have begun to explore the use of LLMs in generating program properties such as weakest preconditions~\cite{king2025fuzzfeedautomaticapproachweakest, 11024288}, postconditions~\cite{endres2024can}, and full program specifications~\cite{ma2024specgen}. Concurrently, researchers have also started to investigate the use of LLMs for generating loop invariants; the works of~\cite{kamath2023finding, pei2023can, janssen2024can} lay the foundations for such investigations and establish the notion that LLMs can indeed generate useful loop invariants. Since then, a plethora of further works has emerged.

The authors of~\cite{pirzada2024llm} generate loop invariant candidates with LLMs; these are then validated by the Vampire theorem prover~\cite{kovacs2013first} and used to replace loops in the context of bounded model checking~\cite{clarke2001bounded}. Similarly, the work of~\cite{wu2024llm} uses both LLMs and bounded model checking: their method employs LLMs to generate loop invariant candidates and bounded model checking to filter incorrect predicates. The Lemur framework~\cite{wu2023lemur} uses LLMs to generate and repair candidate loop invariants, which are then used within the ESBMC~\cite{menezes2024esbmc} and UAutomizer~\cite{heizmann2016ultimate} frameworks for program verification.

A common theme that emerges from these recent works is the benefit of integrating LLMs with formal tools; these approaches can be considered under the umbrella of “neurosymbolic AI”. Our work follows a similar tack, but distinguishes itself through: (1) systematic backward chaining using LLM-based weakest precondition reasoning; (2) dual-level neurosymbolic refinement and repair that combines internal, LLM-based Hoare-logic reasoning with external OpenJML feedback; (3) automatic handling of (sequentially composed) multi-loop programs; and (4) a focus on Java, in contrast to existing works which primarily target C.

\section{Conclusion}
\label{sec:conclusion}
This paper presents NeuroInv, a neurosymbolic approach to loop invariant generation that features dual-level refinement and repair. NeuroInv contains two key modules: (1) a neural reasoning module that leverages LLMs and Hoare logic to derive and refine candidate loop invariants via backward-chaining weakest precondition reasoning, and (2) a verification-guided symbolic module that iteratively repairs invariants using counterexamples from OpenJML.

We evaluate NeuroInv on a comprehensive benchmark of 150 Java programs, this benchmark features programs which include: single loops, multiple (sequential) loops, multiple arrays, random branching, and noisy code segments. NeuroInv achieves a $99.5\%$ success rate, substantially outperforming ad-hoc prompting ($66.4\%$), neural-only ablations ($84.7\%$), and a state-of-the-art LLM-based specification generator for Java ($28.5\%$)~\cite{ma2024specgen}.

Additionally, a hard benchmark of $10$ larger multi-loop programs (containing an average of $7$ loops) further demonstrates that NeuroInv scales to more complex contexts. In this setting, the performance gap between NeuroInv and ad-hoc prompting widens, with NeuroInv achieving a success rate of $62\%$, whilst ad-hoc prompting only achieves $14\%$.

Given the results achieved, NeuroInv constitutes an effective framework for loop invariant generation. The future work we have outlined is intended to extend this neurosymbolic approach to broader and more demanding verification scenarios.



\bibliography{lipics-v2021-sample-article}

\appendix









\section{Prompts}\label{prompts-appendix}
\begin{lstlisting}[language={},caption={Ad-hoc prompt for loop invariant generation.},label={lst:zero-prompt-loop-inv}]
Compute the loop invariant(s) for the loop(s) in the following Java 
code. Please use the format of JML annotations. Annotate the loop(s) 
with the invariant(s). Provide no additional explanations beyond the 
program code and the required annotation. Reason through your 
solution internally.
\end{lstlisting}

\begin{lstlisting}[language={},caption={The NeuroInv prompt for weakest precondition calculations.},label={lst:wp-gen-neuroinv}]
**Task: Calculate Weakest Precondition**

You are given a Java program with segmented code. Your task is to 
calculate the weakest precondition (WP) for a specific code 
segment with respect to a given postcondition.

**Segmented Program:**
```java
{segmented_program}
```

**Code Segment to Analyze:**
Segment index: {segment_index} {tag_description}

**Postcondition (must be true AFTER the segment executes):**
{postcondition}

**Instructions:**
1. Identify the code segment between the specified tags
2. Apply weakest precondition calculus rules to work backward 
through the statements
3. For assignments: WP(x := e, R) = R[x -> e] (substitute e for x 
in R)
4. For sequences: WP(S1; S2, R) = WP(S1, WP(S2, R))
5. For conditionals: WP(if B then S1 else S2, R) = (B => WP(S1, 
R)) AND (NOT B => WP(S2, R))
6. Simplify your result as much as possible
7. Express the result using logical notation 
(AND, OR, =>, NOT, FORALL, EXISTS)

**Example:**
For code segment: `x = x + 1;` with postcondition: `x > 5`
The WP would be: `x + 1 > 5` which simplifies to `x > 4`
\end{lstlisting}

\begin{lstlisting}[language={},caption={The NeuroInv prompt for loop invariant generation.},label={lst:loop-inv-gen-neuroinv}]
**Task: Generate Loop Invariant**

You are given a Java program with segmented code. Your task is to 
generate a suitable loop invariant for a specific while loop.

**Segmented Program:**
```java
{segmented_program}
```

**While Loop to Analyze:**
Loop index: {loop_index} (between tags `// while {loop_index} 
open` and `// while {loop_index} close`)

**Loop Postcondition (must be true after loop exits):**
{loop_postcondition}

**Instructions:**
1. Identify the while loop between the specified tags
2. Extract the loop guard condition B
3. Analyze the loop body to understand what the loop does
4. Generate the weakest possible loop invariant I that:
   - Is true before the first iteration (initialization)
   - Is preserved by each iteration (I AND B => WP(loop_body, I))
   - Combined with loop exit (I AND NOT(B)) establishes the loop 
   postcondition, {loop_postcondition}
5. Express the invariant using logical notation 
(AND, OR, =>, NOT, FORALL, EXISTS)

**A good loop invariant typically includes:**
- Bounds on loop variables (e.g., 0 <= i <= N)
- Relationships between variables (e.g., i < j)
- Quantified properties about arrays/data structures (e.g., 
FORALL(k). 0 <= k < i => a[k] > 0)
- Progress measures (what has been accomplished so far and what 
remains to be accomplished)

Note: when an existential quantifier is needed in the loop 
invariant, perform the following steps:
1. Introduce and initialise a ghost variable before the loop
2. Add a "//@ set" annotation at the end of the loop body to show 
how this variable needs to change
3. Express the invariant in terms of this variable
\end{lstlisting}

\begin{lstlisting}[language={},caption={The NeuroInv prompt for checking implication one.},label={lst:imp-one-check-neuroinv}]
**Task: Verify Implication One (Loop Preservation)**

You are given a loop invariant and must verify that it is 
preserved through one iteration of the loop.

**Segmented Program:**
```java
{segmented_program}
```

**While Loop:**
Loop index: {loop_index} (between tags `// while {loop_index} 
open` and `// while {loop_index} close`)

**Loop Invariant (I):**
{loop_invariant}

**Loop Guard (B):**
{loop_guard}

**Weakest Precondition of Loop Body w.r.t. I:**
WP(loop_body, I) = {wp_loop_body}

**Implication to Verify:**
I AND B => WP(loop_body, I)

**Explanation:**
If the invariant holds and the loop continues (guard is true), 
does executing the loop body preserve the invariant?

**CRITICAL CONSTRAINT:**
A loop invariant must be **self-sufficient** for formal 
verification. You may ONLY use `I AND B` to prove WP(loop_body, I). 
You MUST NOT reference:
- Method preconditions
- Variable values from before the loop
- Any facts not explicitly stated in the invariant I

If you need external facts (like preconditions) to complete the 
proof, the implication **does NOT hold** - the invariant is 
insufficient and must be strengthened.

**Instructions:**
1. Assume ONLY that I and B are true
2. Using ONLY these facts, determine if WP(loop_body, I) 
logically follows
3. Do NOT use preconditions, even if variables are unmodified
4. If you need any additional facts to prove WP(loop_body, I), 
the implication FAILS
5. If it fails, identify exactly what facts are missing from the 
invariant
\end{lstlisting}

\begin{lstlisting}[language={},caption={The NeuroInv prompt for checking implication two.},label={lst:imp-two-check-neuroinv}]
**Task: Verify Implication Two (Loop Termination Correctness)**

You are given a loop invariant and must verify that when the loop 
exits, the desired loop postcondition is satisfied.

**Segmented Program:**
```java
{segmented_program}
```

**While Loop:**
Loop index: {loop_index} (between tags `// while {loop_index} 
open` and `// while {loop_index} close`)

**Loop Invariant (I):**
{loop_invariant}

**Loop Guard (B):**
{loop_guard}

**Loop Postcondition (must be true after loop exits):**
{loop_postcondition}

**Implication to Verify:**
I AND NOT(B) => LoopPostCondition

**Explanation:**
If the invariant holds and the loop exits (guard is false), is 
the loop postcondition satisfied?

**CRITICAL CONSTRAINT:**
A loop invariant must be **self-sufficient** for formal 
verification. You may ONLY use `I AND NOT(B)` to prove the loop 
postcondition. You MUST NOT reference:
- Method preconditions
- Variable values from before the loop
- Any facts not explicitly stated in the invariant I

If you need external facts (like preconditions) to complete the 
proof, the implication **does NOT hold** - the invariant is 
insufficient and must be strengthened.

**Instructions:**
1. Assume ONLY that I and NOT(B) (negation of the loop guard) are 
true
2. Using ONLY these facts, determine if the loop postcondition 
logically follows
3. Do NOT use preconditions, even if variables are unmodified
4. If you need any additional facts to prove the loop 
postcondition, the implication FAILS
5. If it fails, identify exactly what facts are missing from the 
invariant
\end{lstlisting}

\begin{lstlisting}[language={},caption={The NeuroInv prompt for loop invariant refinement.},label={lst:refine-inv-neuroinv}]
**Task: Refine Loop Invariant**

A loop invariant has failed verification. Your task is to refine 
it by strengthening or adjusting it to satisfy the Hoare logic 
requirements.

**Segmented Program:**
```java
{segmented_program}
```

**While Loop:**
Loop index: {loop_index} (between tags `// while {loop_index} 
open` and `// while {loop_index} close`)

**Current Invariant (FAILED):**
{current_invariant}

**Failed Implication:**
{implication_name}

**Failure Reason:**
{failure_reason}

{refinement_guidance}

**Instructions:**
1. Analyze why the current invariant failed
2. Identify what additional conditions or strengthenings are 
needed
3. Propose a refined invariant that:
   - Still holds initially (before the first iteration)
   - Is strong enough to pass the failed implication
   - Is not overly restrictive (can be maintained by the loop)
4. Explain your refinement strategy
\end{lstlisting}

\begin{lstlisting}[language={},caption={Refinement guidance for implication one failures.},label={lst:refine-imp-one-neuroinv}]
**Refinement Strategy for Implication One:**
- The invariant is not preserved through the loop body
- The invariant must be self-sufficient - it cannot rely on 
preconditions or external facts
- Consider strengthening the invariant by adding conditions about:
  - Facts from preconditions about unmodified variables/arrays 
  (if needed to prove preservation)
  - Variables modified in the loop body
  - Relationships that must be maintained across iterations
  - Bounds that are always respected
  - Properties that hold before and after each iteration
- Ensure the strengthened invariant still allows initialization 
(holds before first iteration)
\end{lstlisting}

\begin{lstlisting}[language={},caption={Refinement guidance for implication two failures.},label={lst:refine-imp-two-neuroinv}]
**Refinement Strategy for Implication Two:**
- The invariant + loop exit doesn't imply the loop postcondition
- The invariant must be self-sufficient - it cannot rely on 
preconditions or external facts
- Consider strengthening the invariant to explicitly include:
  - Facts from preconditions about unmodified variables/arrays 
  (e.g., if precondition says a[k] >= k and 'a' is never 
  modified, add this to the invariant)
  - Properties that hold throughout but aren't stated in the 
  current invariant
  - What has been accomplished when the loop exits
  - Relationships needed for the loop postcondition
  - Quantified properties over the full range
- Ensure the strengthened invariant can still be preserved by the 
loop
\end{lstlisting}

\begin{lstlisting}[language={},caption={The NeuroInv prompt for 
loop invariant repair.},label={lst:repair-inv-neuroinv}]
**Task**
For the Java method {method_name} shown below (under **Program**), 
repair ONLY the loop specifications to resolve the 
verification issues identified by OpenJML. 
Specifically:
- You MAY modify loop invariants and, if present, the loop 
variant (decreases clause).
- You MUST NOT modify method preconditions, postconditions, 
assignable clauses, or any other JML specifications outside of 
the loop.
- You MUST NOT modify the Java code.

The OpenJML output (under **OpenJML Output**) describes the 
failures. Use it carefully to strengthen the loop invariant and/
or adjust the variant to ensure inductiveness and termination as 
required.

**Scope**
- Allowed changes:
  - Loop invariants
  - Loop variants (if present)
- Forbidden changes (leave exactly as-is):
  - Method-level requires/ensures/assignable clauses
  - Class-level specifications
  - Any Java implementation code

**Deliverable**
Return the complete program with only the loop invariant(s) (and 
variant(s), if present) updated. Do not include explanations or 
comments beyond what is explicitly requested.

**OpenJML Output**
{openjml_output}

**Program**
{program}
\end{lstlisting}

\end{document}